# Taxing the Internet

## is that feasible ?

**Sowmyan Jegatheesan**
**December 2014**

 Governments across the globe are facing challenging times to generate more revenue because of the economic slowdown and to balance their budgets. There is a growing need to find new ways of revenue generation as spending cuts and austerity measures don't go well with most sections of the society, especially during difficult economic times. Internet Today is seen more as a necessary commodity and nobody can deny the fact that the Internet has improved peoples' life in an unprecedented way than any other technology in the past. The Internet as a commodity is taxed in many countries, but the Internet usage or the transactions carried out are something that's not taxed.

# Introduction

> *"Nothing can be said to be certain, except death and taxes"*
>
> - Benjamin Franklin

In reality we are taxed for everything and there is no way we can escape from taxes on most of the things we use on a day to day basis. But the internet is a medium in which lot of transactions happen and is not taxed for the nature of the transactions that take place. We are not taxed for what type of transaction that we carry out on the Internet or the amount of data or time spent on the internet directly, but Internet as a commodity is taxed the regular taxes that we pay for other services or commodities. If our monthly internet bill is $50 we might be taxed a fixed percentage of Value Added tax, Sales Tax or other taxes based on our tax jurisdiction. We cannot deny the fact that it will be charged for any other product or commodity. But imagine if we pay a special tax when we used the internet 24*7 or used more data or buying products from the internet from an online store in another tax jurisdiction. The world has seen some of the weirdest taxes levied on items since the Egyptian and Roman empires (Efile.com, 2015). Even today we are paying taxes unknowingly on many items which may seem weird like the internet taxes. We pay a few cents extra when we alter a Bagel for breakfast in New York or pay 6% when we get a tattoo in Arkansas but why not the Internet? (Raymond, 2015)

Today, people are becoming more and more dependent on the Internet and the younger generation has an addiction to it. An average American teen Snapchats more than 150 times a day and its increasing since (Blodget, 2013). We have cable Internet, mobile internet, Public WIFIs and many more ways that let people connect to the World Wide Web. So how will the taxing be possible? Any other communication medium like postal system or telephone system did not have the continuous usage like the internet as we called someone in telephone or mailed an envelope only when we needed, but we are connected to the internet - online all the time sending and receiving emails, having a conversation, watching a movie, buying stuff, doing business or transactions and many more things. So the complexity of taxing or even finding the usage becomes complex. Because it has much wider adaptation since it has become a basic necessity like air, water, food or shelter! We have free public WIFIs in airports and in restaurants, etc. Internet taxing might create accessibility hurdles for people.

## Proposals to tax the Internet – Why Governments want to do it

Government needs to make money as most of the developed countries have a deficit in their annual budgets (Cia.gov, 2015). The deficit is increasing year by year and some countries (Ex. India (De, 2012)) have taken a policy decision not to have more than a certain percent of their GDP as a deficit. The world has faced the biggest economic slowdown since the great depression and the first major recession in the Internet era. Governments went into defensive mode when it came to spending and the wanted to save jobs, bail out certain industries, etc. So the underlying idea is to make more money or find sources of more income. Most of the developed world has the highest tax rates. So the things that are not taxed are very less. The corporate world, especially in the United States is paying less in taxes as companies find the so called tax haven countries and limit the exposure to US taxes (Rt.com, 2015).

So the governments are taxing everything and selling all natural resources like air (Spectrum Frequency), water and minerals. Still, as mentioned earlier they have deficits and the costs are increasing year by year. One of the most used services that people use today that's free of any specific usage taxes is the Internet. So the focus of the moment is mainly on how to making tax money from the internet. Nobody owns the Internet and it's not regulated by a single entity, government or a regulatory body. The trade and transactions that happen in the Internet are significant and its increasing exponentially. One simple use case will be the loss of revenue from a government point of view because of the Internet. Consider a simple check deposit transaction that carried out by scanning the image through the Internet and physically depositing the check. There may be many benefits to the bank and the customer for doing an online transaction because the customer gets the convenience and the bank has the satisfaction of customer service. But the government loses at least some tax or revenue. If the customer travels to the bank there will be a means of transport, gas tax, the bank will have to employ someone and the bank pays payroll tax for the bank employee who will service the customer. The online transaction eliminates many of the steps involved which add up to different revenue streams or taxes to the government. Nevertheless, there are expenses too, government provides infrastructure and other benefits; but the main point of contention is there is some revenue less because of the internet. So the governments look for ways to tax the Internet or at least consider debate such a proposal.

## Carriers, ISP, Regulations, Government and Net neutrality

The ongoing debate about regulating the internet, censorship and net neutrality have not found a reasonable solution that's been accepted by governments and scientists. The tax question would certainly add spice to the ongoing debate since the ownership of the internet is debated. The Internet Service Providers in most cases are the telephone companies feel they are the biggest losers since the internet came as not many people use the telephone rather calls are routed to the internet and the carriers have no way to check the kind of traffic as they just transport the bits. Recent discussions between services like Netflix using the internet to gain viewers and carriers who operate as TV-Cable service providers are again losers because people get cheap and high quality service from Netflix though the internet. So the carriers and Netflix are locking horns about that (Brodkin, 2012) . The Internet has alienated the carriers and content providers like websites and services like Netflix are making money. In the whole process the government doesn't get direct financial benefit.

## Some Internet taxes

As we all know we are taxed for different commodities and services by the state and federal governments directly and indirectly. Internet as a service is taxed at the state level in the United States and there is no national tax that's charged. The practice is the taxes are passed on to the customers by the ISPs. So we are taxed already for the Internet, but as like any other service or commodity we use. Some states like Tennessee and Wisconsin treat the Internet as telecommunication services and are subject to telecom taxes. Bit tax is something that has sparked the Internet taxing row again with Hungary being the first country with a draft proposal. This bit tax will simply be based on the volume of data that's transferred. This is exclusive to the Internet and can be called a core Internet tax as bits are the lifeline of data transmission. The Email tax is an odd version of the bit tax, which will be based on the email sent or received volume. The United Nations Development Program of 1999 "Globalization with a Human Face" mentioned this for the first time. As per the 1996 estimates it would have been $70 billion in revenue (Hdr.undp.org, 2015). There is also a bandwidth tax, which would be based on the speed of our Internet connection.

## Hungary and Internet tax - The Spark

Though different countries in the world discussed the possibility of taxing the internet since the 1990s Hungary was the first country to come up with a proposal. The idea of taxing Internet usage was put forward by the Hungarian government as a draft law few months before. The Hungarian government has various taxes, which the citizens of the country are subjected to. As soon as the proposal was out, the people came to the streets to protest the decision. It gained worldwide media attention, especially because it is seen as a test case. The proposal was severely criticized by the European Union as the union was shocked, especially at a time when they are working to remove mobile roaming within the EUEA by the end of 2015 and one of its member states coming with a fancy proposal to tax the internet usage where people using mobile data would be subject to bit tax.

The proposed tax had a levy at 150 Forints ($0.60) per gigabyte of data traffic. This would have fetched an estimated $440 million to the Hungarian government. After the citizens protested the government decided to have a cap at ($2.8)700 Forints per month for individuals and ($20) 5000 Forints for companies. But the citizens continued protests and finally the government had to drop the plan. (BBC News, 2014)

## United States Freedom of Internet taxes Act

In 1998 the US Congress passed the Internet Tax Freedom Act. When the legislation was signed first it was designed in such way to provide a temporary 3 year moratorium which was extended four times till 2014. On July 15, 2014, the US House of Representatives passed the Permanent Internet Tax Freedom Act , a bill that would amend the Internet Tax Freedom Act and remove the temporary moratorium to make permanent the ban on state and local taxation of Internet access and on multiple or discriminatory taxes on electronic commerce (Congress.gov, 2015).

The Permanent Internet tax freedom Act will also remove the ability for seven states (Hawaii, New Mexico, North Dakota, Ohio, South Dakota, Texas, and Wisconsin) to tax internet access because they implemented internet access tax before the 1998 Internet Tax Freedom Act. Some of the key considerations are Internet access and the internet economy, which are driving the economic growth, equality and productivity. According to a 2012 OECD report "Internet is becoming a key economic infrastructure revolutionizing business and serving as a platform for innovation." (OECD Internet Economy Outlook 2012, 2012) The US internet economy is growing consistently and a tax would slow down the growth and access to the Internet would be slowed down. During the tax moratorium since 1998 the internet penetration and access has increased to more than 80% from 20% and not taxing it has certainly helped the free flow and wide adoption. The comments made by Rep.Bob Goodlatte, who sponsored the Permanent Internet tax freedom act stated after passing the bill in house explains the fact "This legislation prevents a surprise tax hike on Americans' critical services this fall…It also maintains unfettered access to one of the most unique gateways to knowledge and engine of self-improvement in all of human history." (Volz, 2015)

## World Wide Tax projections if internet is taxed

Based on some of the widely available Internet usage statistics the amount of tax money that can be collected is calculated here. These are not based on any economic or financial models, but simply based on the methods proposed in the past.

Email tax – As noted before the United Nations Development Program of 1999[8] proposed an email tax. The report proposed tax of 0.01 cents for every email sent. According to a statistics 182.567 Billion - E-mail traffic per day worldwide (Factshunt.com, 2015).This would be $ 182.567million a day just with emails. But there are some practical problems like spamming, for example. The same statistics say about 43.57% emails are spam. The other issue would be jurisdiction. Countries in where the email servers' are present vs. countries from which the emails are sent. So which country gets the right to tax and above all who will be taxed and how the email flow will be controlled in an unregulated open and free internet?

Bit tax – "The UC San Diego – How Much Information" (Short, Bohn and Baru, 2011) report mentions that in 2008 about 9.57 Zetta bytes of data was transmitted across the world. Which were $9.57*10^{21}$ bytes per year in 2008. The internet usage has more than doubled since then. Just to apply the same tax proposal that the Hungarian government wanted to apply – $0.6 per GB this would account to $ 5.742 trillion. Since 2008 the Internet has more than doubled and in today's estimates would almost equal to US GDP which is $16 trillion. This statistic includes all data in the network. There is a more recent Cisco report (Cisco Visual Networking Index:

Forecast and Methodology, 2013 – 2018, 2014) which has a conservative estimate of IP data which is close to 51,168 PB per month in 2013 and which is expected to triple by 2018. This gives the estimated 2013 bit tax at $368.4 Billion. But again how this will be calculated country wide and how the jurisdictions would work, what would happen to the mobile internet traffic and roaming, public internet services etc are all debatable. The practical implementation clearly has so many hurdles that make things clear that the implementation would not be easy and it's not going to be acceptable.

## Conclusion

Every time when something that's free, without regulation is subject to strict regulations and costs would have setbacks. The Internet is one of the few things that has a wide adoption and is less controlled and not regulated. If taxing or metering is done, then the wider adoption would have issues. The growth of the internet was phenomenal in the last decade as least developed countries and developing countries had a far quicker adoption. One of the most interesting studies that support not to tax the internet is the analysis by the Phoenix center for advanced legal and public policy (Ford, 2014) which says that if the Internet tax freedom act was not in place, guaranteeing moratorium on Internet taxes we would have lost at least 20 million connections and would continue to alienate people from the adoption of the Internet. The Internet also has also improved social, economical lives of millions across the world.

The counter argument is Governments already has sales or value added taxes to the internet and why not the data consumption tax? Traditional telephones were subjected to taxes and why not the internet. Taxation on the heavy consumption of the internet could pave way for net neutrality (BloombergView.com, 2014). The content that's pushed at internet users is huge and some of which is auto playing audio and video and other services that consume huge data. If there are progressive taxes to data, then people might use data more efficiently and encourage equal opportunity for all regardless of their bandwidth power.

The Internet economy is huge. The internet has created new jobs and empowered the lives of millions across the globe. It has reached the nook and corner of the world. If taxed the internet might still survive since people gain more than they lose. But the government's point of view of making money by taxing the internet to support the government funding programs would be counterproductive. Internet if it's free for ever would still create jobs and add value to the economy.

[Bibliography]

Sowmyan Jegatheesan is an Information Technology Business Analyst and Project Coordinator with extensive expertise in managing and implementing high-performance technology solutions. He holds a Bachelors degree in Information Technology from Vellore Institute of Technology, India, a Masters degree from the University of Ottawa's E- Business Technologies Program and a Masters candidate in the Liberal Arts Information Management Systems program at Harvard University.